\documentclass{article}

\usepackage[final]{nips_2018}

\usepackage[utf8]{inputenc} 
\usepackage[T1]{fontenc}    
\usepackage{hyperref}       
\usepackage{url}            
\usepackage{booktabs}       
\usepackage{amsfonts}       
\usepackage{nicefrac}       
\usepackage{microtype}      
\usepackage{subfloat}
\usepackage{algorithm}
\usepackage[noend]{algpseudocode}
\usepackage[pdftex]{graphicx}
\usepackage{subfig}
\usepackage{tabularx}

\title{Predicting Variable Types in Dynamically Typed Programming Languages}

\author{
  Abhinav Jangda  \\
  Department of Computer Science\\
  University of Massachusetts, Amherst\\
  Amherst, MA 01002 \\
  \texttt{aabhinav@cs.umass.edu} \\
   \And
  Gaurav Anand \\
  Department of Computer Science\\
  University of Massachusetts, Amherst\\
  Amherst, MA 01002 \\
  \texttt{ganand@cs.umass.edu} \\
}

\begin{document}

\maketitle

\begin{abstract}
Dynamic Programming Languages are quite popular because they increase the programmer's productivity. However, the absence of types in the source code makes the program written in these languages difficult to understand and virtual machines that execute these programs cannot produced optimized code. To overcome this challenge, we develop a technique to predict types of all identifiers including variables, and function return types.

We propose the first implementation of $2^{nd}$ order Inside Outside Recursive Neural Networks with two variants (i) Child-Sum Tree-LSTMs and (ii) N-ary RNNs that can handle large number of tree branching. We predict the types of all the identifiers given the Abstract Syntax Tree by performing just two passes over the tree, bottom-up and top-down, keeping both the content and context representation for all the nodes of the tree. This allows these representations to interact by combining different paths from the parent, siblings and children which is crucial for predicting types. Our best model achieves 44.33\% across 21 classes and top-3 accuracy of 71.5\% on our gathered Python data set from popular Python benchmarks.
\end{abstract}

\section{Introduction}
Dynamic Programming Languages like Python, JavaScript have gained the title of ``most popular languages'' because of increased adoption in recent years. Python which was created to be a replacement for Perl as a better scripting language is now being widely adopted in several communities. With the Django and Flask Web Frameworks, Python has been widely adopted as one of the main languages for Web Development. 
And frameworks like PyTorch, Tensorflow, NumPy, SciPy, has made Python the de-facto choice for Machine Learning programming. 
Another popular dynamic language, JavaScript, which was developed as a language for writing Web Applications is now the major language for writing web and desktop applications. 
With popular frameworks like Node.js, programmers are using JavaScript even for client-side applications. And other dynamic languages like Ruby and Dart are flourishing as well.

A major reason for the popularity of using dynamic languages is the programmers' productivity. Unlike statically typed languages like C++, C, Java, a dynamically typed language allows programmers to write code quickly by not requiring to declare the types of each variable which are determined at run-time based on the values assigned to that variable, thereby increasing programmer productivity. 
However, dynamic type system suffers from two major drawbacks. First, since the variable types are not declared in the source code, the source code becomes difficult to understand and extend. For programmers working on the large code base written in dynamic languages, it is hard to understand the control flow of the program if the types are not available at the compile time. 
Second, the work of performing type-checking and generating optimized code is done by the run-time environment instead of the compiler, resulting in extra overhead. 
To overcome this issue, virtual machines (VMs) like Google V8, Mozilla SpiderMonkey, first generate inefficient machine code. The inefficient machine code is executed initially while the VM acquires information about the types of each variable, then the virtual machine compiles to efficient machine code with the types obtained from profiling. 
But this approach takes time to determine the profiling information and hence, the code has to be executed from the slow/inefficient path.

To mitigate these issues, we present a machine learning technique to determine the types of variables in a given program written in a dynamically typed language. Our system can be used by programmers to determine the type of each variable, thereby improving programmer's productivity for reading and understanding, and by the virtual machine to generate the optimized source code with the predicted type information, thereby giving performance improvement. 

Recent advances in Natural Language Processing involving parse trees ~\cite{inside-outside, tree-lstm, CNN-AST, Recursive-Socher} has shown improved performance with modern neural network architectures. Le and Zuidema ~\cite{inside-outside} present Inside-Outside networks that allow information to flow in both directions, top-down and bottom-up. However, their work is limited to a maximum of two children for each node and further they use simple recurrent units which have a problem to learn long term dependencies because of vanishing and exploding gradients. In our case, an AST might have tens to hundreds of children, and the type information for an AST node can come from the combination of paths involving  its children, parent, and siblings. Hence, we combine a Tree-LSTM with Inside-Outside information flow ~\cite{inside-outside} to capture and allow the type information to flow in an AST. We also introduce two techniques to modify the Abstract Syntax Tree before passing it to the model: (i) Tree Re-structuring analysis and (ii) Identifier Sink Node analysis. We present first such data set containing Python programs and types of each identifier. Our model learns to predict the types by learning the composition functions and embedding representation of keywords and operators. With our model we obtain the top-1 accuracy of 44.4\% over this dataset and top-3 accuracy of 71.5\%. The contributions of this work are:
\begin{enumerate}
    \item A technique to predict types of all identifier using Tree LSTM with Inside-Outside information flow~\cite{inside-outside} (Section~\ref{sec:methodology}).
    \item Two modification techniques over Abstract Syntax Tree to handle large number of children, and improve the accuracy of model: Tree Re-structuring analysis(Section~\ref{sec:tree-node}) and Identifier Sink Node Analysis (Section~\ref{sec:identifier-sink-node}).
    \item First data set of Python programs and corresponding type of each identifier. (Section~\ref{sec:dataset}).
    \item Our best model gives a top-1 accuracy of 44.34\% and top-3 accuracy of 71.5\% (Section~\ref{sec:experiments})
\end{enumerate}

\section{Related Work}
In this section, we describe the previous work to predict some program properties and then describe a few existing analogous problems in the Natural Language Processing domain involving parse trees.

\paragraph{Program Properties Prediction} Alon. et al. \cite{alon} developed a technique to predict program properties for e.g. variable names, and method names in both dynamic and static programming languages.
They have developed a novel representation of Abstract Syntax Tree called AST Path, which traverses the path from variable/leaf node to a non-terminal node representing an expression. This representation encodes the information for the identifier and limits the number of nodes, reducing the AST to a linear chain. They have used Conditional Random Fields (CRF) and word2vec for predicting the variable and method names. Their technique predicts the full type name (including the name of package containing that type) of variables in Java given the types of surrounding nodes and the incomplete type name of that variable. Moreover, they only focus on the types which can be predicted by a global type inference engine. In contrast, our model predicts the types of all identifiers including variables, and function return types. Our model performs this prediction by performing just two passes over the AST. Instead of looking at the path surrounding the variable, our technique uses the information from its parent, children, and siblings. Their technique is not applicable to a dynamically typed language because unlike Java, where the type information is available for most of the identifiers and nodes, in Python, the type information is not available for most of the identifiers other than the constants.

\paragraph{Tree-Based Convolution Neural Network (TB-CNN)} Convolutional Neural Network for sequence modeling was introduced by Yoon Kim~\cite{cnn-sentence} for sentiment classification on a text sequence. The idea of convolutions on Tree-based structure was developed by ~\cite{CNN-AST} with the idea of sliding convolutional kernel combined with dynamic pooling ~\cite{dynamic-pooling} to process the AST to extract structural information of a program. Mou et al.~\cite{CNN-AST} have developed tree-based convolutional neural networks for classifying programs according to the behavior, functionality, complexity, etc., and detecting code snippets of certain patterns like unhealthy code pattern. Similar to image classification, TB-CNN loses the locality information and thus cannot be used for classifying the nodes in a trivial way. 

\paragraph{Recursive Neural Network} In natural language processing sentiment classification is a long-standing task. Initial approaches used bag-of-words, which evolved to sequence modeling using linear LSTM which assumes a right-branching sentence. Socher et al.~\cite{Recursive-Socher} have introduced Recursive Neural Network for classifying the sentiment of all nodes in the parse tree of a sentence. They represent the sentence using word vectors and use the parse tree of the sentence to compute the vector representation of all the internal nodes in the tree using the same composition function. Both the composition function parameters along with vector representation of words are learned. The internal representation of the nodes are computed in a bottom-up manner and the node representation incorporates the information of the sub-tree rooted at that node for the purpose of classification. In our case some leaf nodes have to be classified based on the combination of paths that need not be uni-directional.

\paragraph{Tree-LSTM}
The Tree-Structured LSTM introduced in \cite{tree-lstm} had two main contributions: (i) a generalization of LSTMs to tree-structured network topologies, and (ii) Child-Sum Trees to handle arbitrary number of children in the tree nodes. It also replaced simple recurrent units in recursive network with LSTM cells to overcome the problem of vanishing or exploding gradients, making it easy to learn long-distance correlations.
The Child-Sum Tree-LSTM transition equations are the following:
\begin{align}
\tilde{h}_p &= \sum_{k \in C(p)} h_k, \label{eq:treelstm-first} \\
i_p &=\sigma \left( W^{(i)} x_p + U^{(i)} \tilde{h}_p + b^{(i)} \right), \\
f_{pk} &= \sigma\left( W^{(f)} x_p + U^{(f)} h_k + b^{(f)} \right), \label{eq:treelstm-f}\\
o_p &= \sigma \left( W^{(o)} x_p + U^{(o)} \tilde{h}_p  + b^{(o)} \right), \\
u_p &= \tanh\left( W^{(u)} x_p + U^{(u)} \tilde{h}_p  + b^{(u)} \right), \\
c_p &= i_p \odot u_p + \sum_{k\in C(p)} f_{pk} \odot c_{k}, \\
h_p &= o_p \odot \tanh(c_p), \label{eq:treelstm-last}
\end{align}
Here $C(p)$ represents the children of node $j$ and $h_p$ and $c_p$ are the hidden and cell state of $p^{th}$ node. We combine this extension with the Inside-Outside algorithm.

\begin{figure}[h]
\centering
\begin{tabularx}{\textwidth}{cc}
\subfloat[Inside-Outside Recursive Neural Network (IORNN). Black rectangles correspond to inner representations, white rectangles correspond to outer representations.]{
\includegraphics[width=0.45\linewidth]{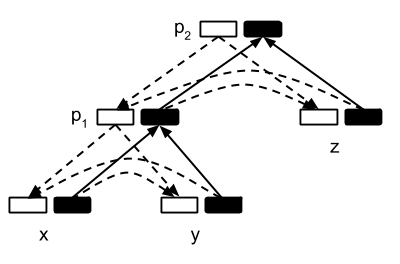}
} &
\subfloat[Inner ($i_p$) and outer ($o_p$) representations
at the node that covers constituent p. They are vectorial representations of p’s content and context,
respectively.] {
\includegraphics[width=0.45\linewidth]{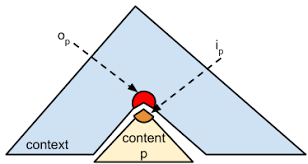}
}
\end{tabularx}
\caption{}
\label{iornn-img}
\end{figure}

\paragraph{Inside-Outside Recursive Neural Network} The architecture and algorithm introduced by Le and Zuidema~\cite{inside-outside} allows information to flow not only bottom-up, as in traditional recursive neural network, but top-down as well. Every node in the hierarchical structure is associated with two vectors: (i) inside representation to represent content under the node, and (ii) outside representation to represent context (see Figure~\ref{iornn-img}). The inside vectors are computed in the bottom-up (post-order traversal) manner combining the inside representation of the children, thus capturing the content of the sub-tree. The outside vectors are computed in the top-down (pre-order traversal) manner combining the representation of the parent's outside (context) and the siblings' inside (content), thus capturing the rest of the tree (see Figure~\ref{iornn-img}). This approach is very critical when we want to classify leaf nodes and the information may come from any combination of paths in the tree.

\section{Methodology}
\label{sec:methodology}


An abstract syntax tree (AST) represents syntactic organization of the source code for a language. Syntax Analyzer also called parser is used to convert the source code into an Abstract Syntax Tree (AST). In our technique, we target Python source code. We use Python's built-in \texttt{ast} module to convert the source code to Python's Abstract Syntax Tree. Our model is based on the work of~\cite{inside-outside}.  Figure~\ref{fig:python-ex} shows a simple Python program, where types of all identifiers is \texttt{int}. The addition operator $+$ is overloaded in Python for \texttt{int}, \texttt{float}, \texttt{string} and custom types including classes. Hence, according to the semantics of $+$ operator, addition of two integers is always integer. Figure~\ref{fig:python-ast-ex} shows AST for Python program in Figure~\ref{fig:python-ex}. Our techniques takes the AST of the program as the input and returns the types of all identifier nodes in the AST.

\newsavebox\pythonexbox
\begin{lrbox}{\pythonexbox}
\begin{lstlisting}
a = 1
b = 2
c = a + b
\end{lstlisting}
\end{lrbox}

\lstset{language=Python}
\begin{figure}
\begin{tabularx}{\textwidth}{cc}
\subfloat[Python program with types for \texttt{a},  \texttt{b}, and \texttt{c} as \texttt{int}]{
\usebox \pythonexbox
\label{fig:python-ex}
}\hspace{10em}

&
\subfloat[Python AST for the program in Figure~\ref{fig:python-ex}. \texttt{a}, \texttt{b}, and \texttt{c} represents the identifiers for which the types have to be predicted.]{
\includegraphics[width=0.45\linewidth]{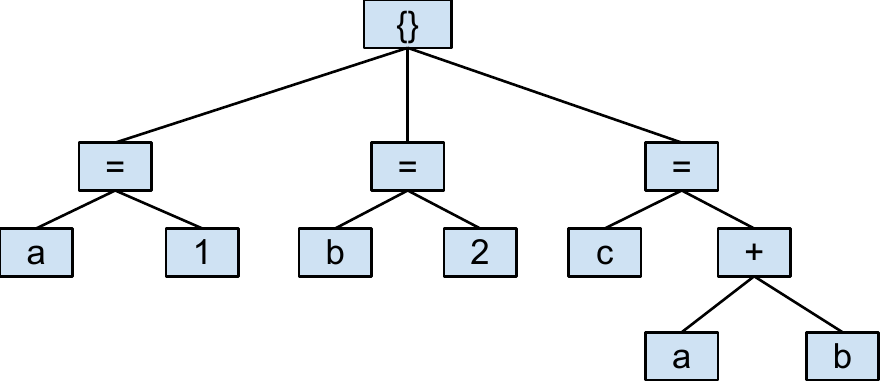}
\label{fig:python-ast-ex}
}
\end{tabularx}
\end{figure}

\subsection{Architecture}
All the the keywords (including operators) in Python are stacked in the word embedding matrix L $\in \mathbb{R}^{D_i}$ of size $(\lvert V \rvert+2) \times D_i$, where $\lvert V \rvert$ is the size of vocabulary and $D_i$ is the size of vector embeddings. The two extra words are, \emph{VAR} and \emph{UNK}, which are used for identifiers and unknown tokens respectively. 

\subsubsection{Node/Class Labels} Our technique works on 21 Python built-in types including common types like:
\begin{enumerate*}
    \item \texttt{int}
    \item \texttt{string}
    \item \texttt{float}
    \item \texttt{None}
    \item \texttt{module}
    \item \texttt{tuple}
    \item \texttt{dict}
    \item \texttt{list}
    \item \texttt{set}
\end{enumerate*}
For each identifier we associate a label which is the type of that identifier in the program. Hence, each of the above types are the labels of the identifiers. With a random model we will get an accuracy of $4.76\%$. To help learning for the model we also add built-in functions of Python and their return type. These built-in functions are:
\begin{enumerate*}
    \item \texttt{int}
    \item \texttt{string}
    \item \texttt{float}
    \item \texttt{tuple}
    \item \texttt{dict}
    \item \texttt{list}
    \item \texttt{set}
    \item \texttt{range}
    \item \texttt{xrange}
    \item \texttt{len}
    \item \texttt{reversed}
\end{enumerate*}

\subsubsection{Tree Nodes}
\label{sec:tree-node}
Our architecture takes a modified AST as input. We convert the given Python AST into a tree format where each node contains: (i) the corresponding Python AST node for e.g. \texttt{while}, \texttt{/}, \texttt{*} etc., (ii) all children nodes, (iii) the inside and outside vectors. Each tree node has its own embedding index to index a row in the embedding matrix. For keyword nodes we return an embedding index from 0 to $\lvert V \rvert - 1$. For identifier nodes we return the embedding index as $\lvert V \rvert + 1$ and all other keywords are returned with embedding index of $\lvert V \rvert$.

\subsection{Models}
The extension of keeping inside and outside vectors is same for both the composition function variants that we tried. The outside vector summarizes all the information about the context of node p, and inside vector recapitulates the sub-tree at the node p.

\subsubsection{Inside Outside Child-Sum Tree-LSTM}
The Child-Sum Tree-LSTM unit conditions its components on the sum of child hidden states $h_k$, thus it is suitable for trees with high-branching factor or whose children are unordered.

\paragraph{Computing Inside Representation}
The inner representations are computed bottom-up. For internal node p, the input is the embedding of the keyword $x_p \in \mathbb{R}^{D_i}$, and the hidden and cell state, $h_p, c_p \in \mathbb{R}^{D_m}$  for inner representation use the same equations as the ones shown for Child-Sum tree LSTM \ref{eq:treelstm-first} to \ref{eq:treelstm-last}.

\paragraph{Computing Outside Representation}
The outer representations are computed top-down for node p, except the root node for which the inside representation is equal to the outside representation. The outer representation uses the following equations: 
\begin{align}
\tilde{h}_p &= \sum_{k \in S(p)} h_k, \label{eq:childtreelstm-first} \\
i_p &=\sigma \left( U^{(i)} \tilde{h}_p + b^{(i)} \right), \\
f_{pk} &= \sigma\left( U^{(f)} h_k + b^{(f)} \right), \label{eq:childtreelstm-f}\\
o_p &= \sigma \left( U^{(o)} \tilde{h}_p  + b^{(o)} \right), \\
u_p &= \tanh\left(  U^{(u)} \tilde{h}_p  + b^{(u)} \right), \\
c_p &= i_p \odot u_p + \sum_{k\in S(p)} f_{pk} \odot c_{k}, \\
h_p &= o_p \odot \tanh(c_p), \label{eq:childtreelstm-last}
\end{align}
Here $S(p)$ represents the concatenated list of parent and siblings of node $p$. $h_p$ and $c_p$ are the hidden and cell state of $p^{th}$ node. 
\subsubsection{Inside Outside N-ary RNN}
The N-ary RNN with simple recurrent units is suitable where the branching factor is at most \maxchildren children and the children are ordered. Although this requires parameters proportional to the maximum number of children, but it allows the model to learn more fine-grained conditioning on the states of a unit's children when compared to the Child-Sum Tree-LSTM.
\paragraph{Computing Inside Representation}
The inner representations are computed bottom-up. For every node $p$, the input contains: (i) the embedding of the keyword $x_p \in \mathbb{R}^{D_i}$, and (ii) the hidden and cell state, $h_p, c_p \in \mathbb{R}^{D_m}$, for the inner representation use the following equations:
\begin{align}
    h_{p}^i = \tanh(W^i x_p + \sum_{k \in C(p)} U_k^{i_h} h_{k}^i) \\
    c_{p}^i = \tanh(W^i x_p + \sum_{k \in C(p)} U_k^{i_c} c_{k}^i)
\end{align}

\paragraph{Computing Outside Representation}
For the outer representations there is no input and they are computed in top-down order for node $p$, except in the case of the root node for which the inside representation is equal to outside representation. The outer representation is computed using the following equations: 
\begin{align}
    h_{p}^o = \tanh(W^o h_p^o + \sum_{k \in S(p)} U_k^{o_h} h_{k}^o) \\
    c_{p}^o = \tanh(W^o c_p^o + \sum_{k \in S(p)} U_k^{o_c} c_{k}^o)
\end{align}
Here $S(p)$ represents all siblings of node $p$. $h_p^o$ and $c_p^o$ are the hidden and cell state of the outside representation of $p^{th}$ node respectively, and $h_k^o$ and $c_k^o$ are the hidden and cell state of the outside representation of the $k^{th}$ sibling of node $p$ respectively.


\subsection{Algorithm}

\begin{algorithm}[th]
\caption{Type Prediction Model}
\label{algo:type-pred-model}
\small
\begin{algorithmic}[1]
\Function{TypePrediction-Training}{astDataSet} 
\State internalTreeDataSet $\gets$ convertASTToTree(astDataSet) \label{line:train:astToInternalTree}
\State addNameSinkNodes (internalTreeDataSet)\label{line:train:sinkNodes}

\State treeReStructuring (internalTreeDataSet)\label{line:train:treeRestructuring}
\State model $\gets$ Model () \label{line:train:initModel}
\For {\textbf{each} {epoch}} \label{line:train:begin}
    \For {tree, labels $\in$ internalTreeDataSet} 
        \State logits $\gets$ ForwardPass(model, tree) \label{line:train:forwardpass}
        \State loss = CrossEntropyLoss(logits, labels)
        \State loss.backward()
    \EndFor
\EndFor\label{line:train:end}
\Return model
\EndFunction
\end{algorithmic}
\end{algorithm}

\begin{algorithm}[th]
\caption{Forward Pass}
\begin{algorithmic}[1]
\Function{ForwardPass}{model, tree}
\For {node $\in$ tree.postorder ()} \label{line:forward:bottom-up}
    \State {node.inside\_h, node.inside\_c $\gets$ model.apply\_inside\_computation (node)}
\EndFor
\State {tree.root.outside\_h $\gets$ tree.root.inside\_h}
\State {tree.root.outside\_c $\gets$ tree.root.inside\_c}
\For {node, parent, siblings $\in$ tree.preorder ()} \label{line:forward:top-down}
    \State {node.outside\_h, node.outside\_c $\gets$ model.apply\_outside\_computation (node, parent, siblings)}
    \\
    \Return {ReLU(LinearTransform(Unknown\_Vars\_Outside\_h)}
\EndFor
\EndFunction
\end{algorithmic}
\end{algorithm}

Function \textsc{TypePrediction-Training} in Algorithm~\ref{algo:type-pred-model} is the training procedure of our model. It takes the data set of ASTs and returns the trained model after performing Stochastic Gradient Descent using Adam optimizer. Line~\ref{line:train:astToInternalTree} first converts the given AST data set into an internal tree representation~\ref{sec:tree-node}. Line~\ref{line:train:sinkNodes} then performs our Identifier Sink Node Analysis (Section~\ref{sec:identifier-sink-node}) to add sink nodes to the identifier nodes that  represents the same variables in the scope. Line~\ref{line:train:treeRestructuring} performs our Tree Restructuring Analysis (Section~\ref{sec:tree-restructuring}) for restructuring the tree such that if a tree node has more than \maxchildren children, then it will recursively add more nodes to limit the number of children to \maxchildren. This analysis helps in keeping the number of children for each node low and still makes sure that all children are considered in training and prediction processes. Line~\ref{line:train:begin}$-$~\ref{line:train:end} performs the training process. Line~\ref{line:train:forwardpass} performs the forward pass on the model by calling \textsc{ForwardPass} function.

Function \textsc{ForwardPass} performs bottom-up and top-down pass to compute inner and outer representations. It first computes the inner representation for every node in the bottom-up (post-order traversal) manner at line~\ref{line:forward:bottom-up} using inner representations of the children. And then computes outer representation in the top-down (pre-order traversal) manner at line~\ref{line:forward:top-down} using the outer representation of parent and inner representations of the siblings thus giving full context. And finally it predicts the types of all unknown leaf nodes using the outside representation which has the full context. Following sections provide the details of each of the above passes.
\begin{figure}
\begin{tabularx}{\textwidth}{cc}
\subfloat[Sink Nodes for \texttt{a} and \texttt{b}]{
\includegraphics[width=0.45\linewidth]{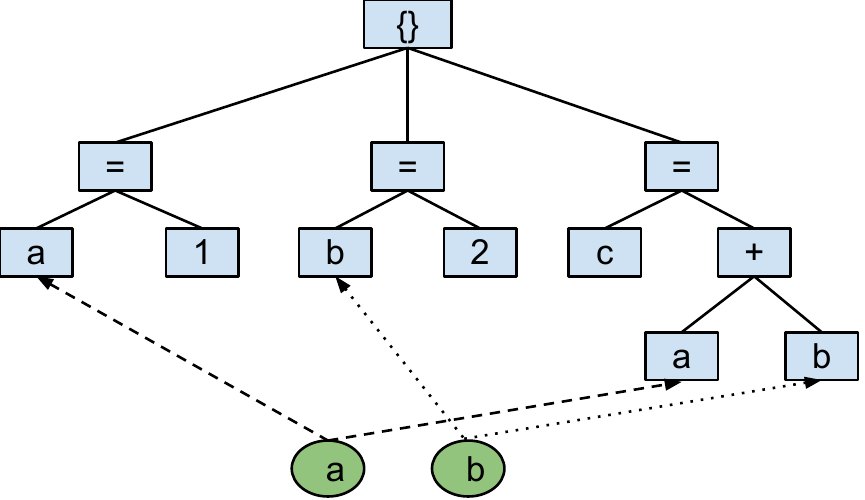}
\label{fig:python-ast-sinknode}
}
&
\subfloat[After Tree Restructuring first two statements are added in an \texttt{if} block, when \maxchildren is 2]{
\includegraphics[width=0.45\linewidth]{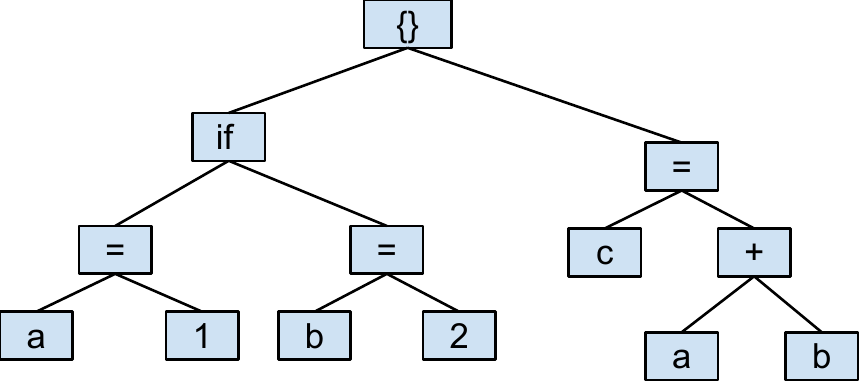}
\label{fig:python-ast-tree-restructuring}
}
\end{tabularx}
\caption{}
\end{figure}
\subsubsection{Identifier Sink Node Analysis}
\label{sec:identifier-sink-node}
In a dynamically typed programming language like Python, each variable can be declared at multiple locations (in the same scope) and can be used (or referenced) at multiple locations. Moreover, each of the declaration and usages are not connected to each other. For instance in Figure~\ref{fig:python-ex} variables \emph{a} and \emph{b} have been defined and used in different statements. As can be seen in the AST of the program in Figure~\ref{fig:python-ast-ex} the use and declaration of variables are not connected with each other. The types are determined when the variable is defined by assigning it a value, for example, the type of \texttt{a} is determined when integer 1 is assigned to it. Unfortunately, there is no type information flow available in the AST that can transfer types the statement where the variable was declared to the statement where the is referenced. Hence, the model will not be able to transfer the type information.

To provide such a flow of type information we, perform \textit{static analysis} to add a new \emph{sink} node which are common for each identifier. Such a node connects the declaration of the variable and all its uses. For example, Figure~\ref{fig:python-ast-sinknode} shows the sink nodes created for variables \emph{a} and \emph{b} which are connecting the declaration of \emph{a} and \emph{b} with their uses. Hence, our model is not predicting the type of identifier nodes but it will now predict the types of \emph{sink} nodes. Once the type of a \emph{sink} node is predicted, it will then transfer the type to other connecting identifier nodes. For example, in Figure~\ref{fig:python-ast-sinknode} when the type information of \emph{a} and \emph{b} are determined then they are passed to the next statement to determine the type of \emph{c} as shown by the arrows.

\subsubsection{Tree Restructuring}
\label{sec:tree-restructuring}
In our model we limit the number of children that each node can get information from (or transfer information to). Such a limitation is required to make sure that the model can learn better from the amount of information that is given and the timing and memory requirement during training and prediction are within acceptable bounds. Unfortunately, large programs can easily have hundred children for each parent. These cases usually arise when each block contains hundred of statements. We can restructure the tree in these cases but we have to make sure that after such a restructuring the program's syntax and semantics are both preserved. A tree restructuring will be invalid if the restructuring changes the types of the identifiers that are to be predicted. In the following text, we denote this limit of number of children as \maxchildren

We perform a tree restructuring analysis based on the insights received from our dataset that each program's AST has hundreds of children only in a block, for example, the statement block of the function's definition can have several statements. Hence, we have to perform tree restructuring only when a block of statements has more than the maximum number of children as the statements. In this case, we split the child into several child blocks, with each block having at maximum \maxchildren number of children. If the new child block also has more than \maxchildren, then we further split these blocks into more blocks. We perform this recursively, by adding more child blocks until each of the node is having at maximum \maxchildren number of children. Since, in Python the \texttt{if} statement block does not add another scope, we use \texttt{if (True)} as our child block. Using \texttt{if (True):} as child block ensures that we do not define variables in a new scope and the block is always executed, thereby, preserving the semantics and syntax of the program.

Figure~\ref{fig:python-ast-tree-restructuring} shows the example of our Tree Restructuring analysis done on the AST in Figure~\ref{fig:python-ast-ex} with \maxchildren set to 2. Our analysis groups the first two statements defining \texttt{a} and \texttt{b} in one \texttt{if (True):} node. Since, \texttt{if} statements does not introduce a new scope, hence, the variables defined in that scope can still be referenced in the block after this \texttt{if} statement. Note that, our analysis however, do not add a new \texttt{if (True):} node for the third statement, we try to add minimum number of extra child blocks.

Since, our model is executed on each node and each node takes information from all its children. The information of each inserted child block will be transferred to the parent node. The information of each child block is determined using the information of each of the statements that were in the original AST. Hence, our model not only limits the number of children but also makes sure that the information is propagated. In Section~\ref{sec:eval-tree-restructuring} we show that using our tree restructuring analysis improves the accuracy.

\section{Data Sets}
\label{sec:dataset}
We use Python benchmarks in~\cite{pypy-benchmarks} as our data set of Python programs. It is a popular set of Python benchmarks which are used to evaluate different implementations of Python. Hence, these benchmarks serve as a good source of data set. 

Variables in a Python program can be defined as (i) local variables in the function, or (ii) global variables. Hence, we need to get types of all local and global variables. We target 21 Python types including all the built-in types like: \texttt{int}, \texttt{string}, \texttt{float}, \texttt{list}, \texttt{tuple}, \texttt{function}, \texttt{module}, and others.  Python provides \texttt{locals} and \texttt{globals} functions, which returns a dictionary of local and global variable names with their value. We first parse all the benchmark files including the standard library used by the Python benchmarks and add extra code for printing values in \texttt{locals} dictionary, by plugging this ast in each function's ast before every return statement and at the end of the function's ast. Moreover, we also add code for printing \texttt{globals} dictionary at the end of each file, to print global variables and their types. After this pass, the code is printed using \texttt{codegen} module. We then execute each benchmark to print the types of local and global variables.

With this approach we retrieve a data set of 23014 number of total AST nodes, containing 1581 identifiers, which are distributed over 36 Python files. In this dataset we obtain 21 class types.

\section{Experiments and Results}
\label{sec:experiments}
In this section we mention in detail about the experiments performed to evaluate and study our approach. Our implementation is developed in Python 3.6 and uses PyTorch framework.

\subsection{Cross-Validation and Hyperparameter Tuning}
We divide our data set into two parts, training set contains 24 ASTs and test set contains 12 ASTs. We perform 4-fold cross-validation, by carving out a validation set.

\paragraph{Hyperparameter Tuning} Our models were trained using Adam optimizer with a learning rate of 0.01 and L2 regularization strength of $10^{-5}$. We select two parameters: (i) the input embedding vector size $D_i$, and (ii) the hidden and cell state vector dimension $D_m$. Table~\ref{tab:hyperparam-accuracy} shows the average accuracy of all folds with different values of the hyperparameters. IO Child-Sum Tree-LSTM performs best when $D_i$ = 10, and $D_m$ = 15. IO N-ary RNN performs best when $D_i$ = 10, and $D_m$ = 10. We pick these hyperparameters for both our models, for the rest of the evaluation. 
\begin{table}
    \centering
    \begin{tabular}{|c|c|c|}
    \hline
         $D_i$ , $D_m$ & IO Child-Sum Tree-LSTM  & IO N-ary RNN\\\hline
         5, 10 & 36.92\% & 35.7\%\\ \hline
         10, 10 & 34.42\% & \textbf{43.0}\%\\ \hline
         10, 15 & \textbf{39.55}\% & 40.6\%\\ \hline
         10, 20 & 39.25\% & 40.3\%\\ \hline
    \end{tabular}
    \caption{The table shows the average accuracy on 4-fold Cross-Validation set of both the models with different values of Dimensions $D_i$ and $D_m$}
    \label{tab:hyperparam-accuracy}
\end{table}
\subsection {Accuracy over test data}
We compared the accuracies of our models with best performing parameters with the accuracy of a random and majority predictor. Table~\ref{tab:test-accuracy} shows the accuracy on test dataset for: (i) IO Child-Sum TreeLSTM with $D_i$ = 10 and $D_m$ = 15, (ii) IO N-ary RNN with $D_i$ = 10 and $D_m$ = 10, (iii) a random predictor to randomly predict one of the 21 classes, and (iv) a majority predictor which predicts the maximum occurring class. 

\subsection{Effect of Max Children (\maxchildren)}
\begin{figure}[H]
    \centering
    \includegraphics[width=0.7\linewidth]{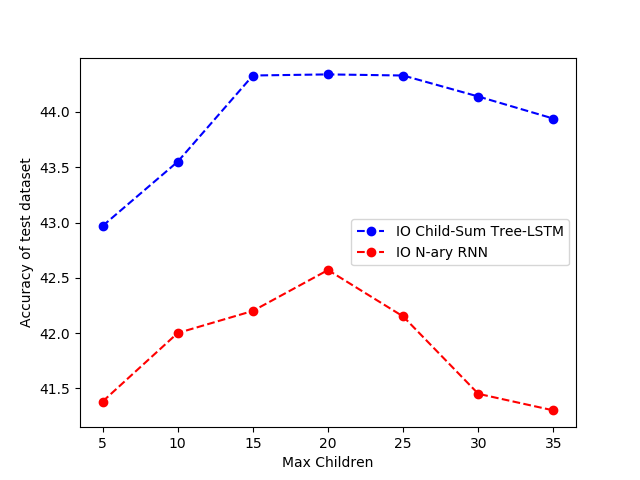}
    \caption{Accuracy on test data vs Max-Children for both models with Re-structuring and best hyperparameters}
    \label{fig:max-children-accuracy}
\end{figure}

We studied the effect of \maxchildren on the accuracy of both the models. For this experiment we use both models with best performing parameters, i.e. IO Child-Sum Tree-LSTM with $D_i$ = 10 and $D_m$ = 15, and IO N-ary RNN with $D_i$ = 10 and $D_m$ = 10. Figure~\ref{fig:max-children-accuracy} shows the accuracies obtained for both models on test data as we increase the value of \maxchildren. 

We can see that at \maxchildren = 20 both models obtain the maximum accuracy. Moreover, the accuracy increases with increase in the value of \maxchildren until 20 and then decreases continuously. 

\begin{table}
\centering
\begin{tabular}{|c|c|c|c|}
\hline
IO Child-Sum Tree-LSTM & IO N-ary RNN & Majority Predictor Baseline & Random Predictor Baseline\\\hline
44.40\% & 42.57\% & 24\% & 4.76\%\\ \hline
\end{tabular}
\caption{Accuracy of both models with best performing hyperparameters on Test set}
\label{tab:test-accuracy}
\end{table}

\subsection{Accuracy of Tree Restructuring}
\label{sec:eval-tree-restructuring}
We perform experiments on our best model to see the improvement in accuracy on test dataset obtained by using Tree Restructuring Analysis. Without tree restructuring, for every node, we only use \maxchildren children and do not take into account other children for IO N-ary RNN. For this experiment we use both models with best performing hyperparameters and best performing \maxchildren, i.e., for IO Child-Sum Tree-LSTM $D_i$ = 10, $D_m$ = 15, and \maxchildren = 20, and IO N-ary RNN with $D_i$ = 10, $D_m$ = 10, and \maxchildren = 20.  Figure~\ref{fig:max-children-accuracy} shows the accuracy on test data with it. 
\begin{table}[H]
    \centering
    \begin{tabular}{|c|cc|cc|}
        \hline
        \specialcell{Max-Children\\ \maxchildren} &  \multicolumn{2}{c|}{IO Child-Sum Tree-LSTM} & \multicolumn{2}{c|}{IO N-ary RNN}\\
        \cline{2-3} \cline{4-5}
         & \specialcell {With \\ Restructuring} & \specialcell {Without\\ Restructuring} & \specialcell{With \\  Restructuring} & \specialcell{Without\\ Restructuring} \\ \hline
         5 & \textbf{42.96} & 41.2 & \textbf{41.38} & 40.5\\\hline
         10 & \textbf{43.55} & 42.12 & \textbf{42.00} & 40.7\\\hline
         15 & \textbf{44.34} & 42.5 & \textbf{42.2} & 41.1\\\hline
         20 & \textbf{44.33} & 42.6 & \textbf{42.57} & 41.4\\\hline
         25 & \textbf{44.14} & 42.45 & \textbf{42.15} & 41.05\\\hline
         30 & \textbf{43.94} & 42.1 & \textbf{41.45} & 40.7\\\hline
         35 & \textbf{43.94} & 41.9 & \textbf{41.3} & 39.1\\ \hline
    \end{tabular}
    \caption{Accuracy on test data set with and without Tree Restructuring for the best performing hyperparameters}
    \label{tab:tree-restructuring}
\end{table}

Results shows that for both models the accuracy with tree restructuring is consistently better than the accuracy without tree restructuring. This confirms our hypothesis that tree restructuring is better even for IO Child-Sum Tree-LSTM.

\subsection{Top-k accuracy}
Since, our model finds the probability of type of each variable. We also perform experiments to determine if the accurate type comes in top 5 types or not. Figure~\ref{tab:top-k-accuracy} shows the results of this experiments.
We can see that there has been significant increase in accuracy from Top 1 to Top 2 and Top 3, however, after that there is not much gains in accuracy. Hence, the developers using our technique can only look at the top 3 predicted types and there will be 71\% chances that it will be the correct. Moreover, virtual machine will need to generate optimized code for only top 3 predicted types.
\begin{table}[H]
\centering
\begin{tabular}{|c|c|c|c|c|c|}
\hline
Model&Top-1 & Top-2 & Top-3 & Top-4 & Top-5\\\hline
IO Child-Sum Tree-LSTM & 44.33\% & 61.33\% & 71.5\% & 78.32\% & 82.42\%\\ \hline
IO N-ary RNN & 42.54\% & 58.8\% & 67.7\% & 73.4\% & 78.1\%\\ \hline
\end{tabular}
\caption{Top 1 to 5 accuracy for both models with best hyperparameters and Tree Restructuring}
\label{tab:top-k-accuracy}
\end{table}

\section{Discussion and Conclusions}
In this work, we have developed a technique to predict the identifiers' type for a  given Python program and/or Python AST. In an AST, the type information for a node can come from the combination of paths involving  its children, parent, and siblings. Hence, our problem is to classify types of some nodes in the tree using the context of the rest of the tree and content of the sub-tree, which in this case is the combination of paths from siblings, parent and children of the node. We have conducted and shown the results on two proposed variants of Inside-Outside networks (i) Inside-Outside Child-Sum LSTM and (ii) Inside-Outside N-ary RNN. We have also shown that re-structuring the tree by splitting children and augmenting parent nodes for N-ary RNN improves the results. Out of the two variants Child-Sum Tree-LSTM performs slightly better. Child-sum LSTM is order-insensitive while N-ary RNN is order-sensitive but requires parameters proportional to the number of children. With our technique we achieve top-1 accuracy of 44.34\% our test data and top-3 accuracy of 71.5\%.

\subsection{Future Work}
In future one should definitely work on expanding the data set so that variants such as N-ary Tree-LSTM can be explored without overfitting as it requires four times more parameters than IO N-ary RNN.
Our technique only focuses on predicting Python's built-in types. Although these are 21 types but these does not contain the types declared by programmers including classes and structs. So far, we set each class instance to an object. Our technique also does not take into account the source code of modules, which are used in Python program. In future work, we will like to support adding information from each module. 

\medskip

\bibliographystyle{plain}
\bibliography{ref}

\end{document}